# ST2HE: A Cross-Platform Framework for Virtual Histology and Annotation of High-Resolution Spatial Transcriptomics Data

Zhentao Liu[1], Arun Das[2,3], Wen Meng[2,4], Yu-Chiao Chiu[3,5,6], Shou-Jiang Gao[2,4], and Yufei Huang[1,2,3,7]✉

[1]Department of Electrical and Computer Engineering, Swanson School of Engineering, University of Pittsburgh, Pittsburgh, PA, USA

[2]Cancer Virology Program, UPMC Hillman Cancer Center, Pittsburgh, PA, USA

[3]Department of Medicine, University of Pittsburgh School of Medicine, Pittsburgh, PA, USA

[4]Department of Microbiology and Molecular Genetics, University of Pittsburgh School of Medicine, Pittsburgh, PA, USA

[5]Cancer Therapeutics Program, University of Pittsburgh Medical Center Hillman Cancer Center, Pittsburgh, PA, USA

[6]Pittsburgh Liver Research Center, University of Pittsburgh Medical Center and University of Pittsburgh School of Medicine, Pittsburgh, PA, USA

[7]Department of Pharmaceutical Sciences, University of Pittsburgh School of Medicine, Pittsburgh, PA, USA.

✉ Corresponding Author

Yufei Huang - email: yuh119@pitt.edu

## Abstract

High-resolution spatial transcriptomics (HR-ST) technologies offer unprecedented insights into tissue architecture but lack standardized frameworks for histological annotation. We present

ST2HE, a cross-platform generative framework that synthesizes virtual hematoxylin and eosin (H&E) images directly from HR-ST data. ST2HE integrates nuclei morphology and spatial transcript coordinates using a one-step diffusion model, enabling histologically faithful image generation across diverse tissue types and HR-ST platforms. Conditional and tissue-independent variants support both known and novel tissue contexts. Evaluations on breast cancer, non-small cell lung cancer, and Kaposi's sarcoma demonstrate ST2HE's ability to preserve morphological features and support downstream annotations of tissue histology and phenotype classification. Ablation studies reveal that larger context windows, balanced loss functions, and multi-colored transcript visualization enhance image fidelity. ST2HE bridges molecular and histological domains, enabling interpretable, scalable annotation of HR-ST data and advancing computational pathology.

**Main**

Spatial transcriptomics (ST) technologies are rapidly advancing, enabling gene expression mapping with preserved spatial context. High-resolution ST (HR-ST) techniques such as Xenium[1], NanoString CosMx SMI[2], and MERFISH[3] allow precise localization of transcripts across diverse tissue types. Despite these advances, no standardized system exists for annotating spatial structures and phenotypes in HR-ST data. Current strategies fall into two categories: (1) Manual expert annotation, which is accurate but subjective and difficult to scale. (2) Computational approaches that cluster cells or analyze neighborhood composition but lack biologically meaningful labels[4–7], as no reference database links spatial transcriptomic patterns to known tissue structures. These limitations highlight the need for frameworks that integrate HR-ST data with high-quality histopathological annotations.

A central challenge is the absence of a reference linking gene expression patterns to histopathological structures. Subsequently, annotations such as tumor subtypes (e.g. ductal carcinoma in situ (DCIS)) cannot be reliably annotated without expert pathological examination (Fig. 1a). Thus, despite HR-ST's resolution, its interpretability remains constrained (Fig. 1a). By contrast, hematoxylin and eosin (H&E) staining remains the gold standard in histopathological examination[8], supported by decades of practice and rich literature[8–10]. Numerous foundation models trained on H&E images including convolutional neural networks[10], transformer-based architectures[11], and CLIP-based models[12–14] capture morphological and pathological features and have proven effective in tumor classification, tissue segmentation, and morphology recognition. Transferring this knowledge to HR-ST could significantly enhance tissue characterization by integrating molecular and histopathological context.

However, paring H&E with HR-ST remains challenging. While Xenium supports same-slide H&E, tissue processing can damage the tissue, thus reducing fidelity. Moreover, researchers must choose between performing H&E or immunofluorescence staining for downstream validation, as conducting one typically precludes the other due to tissue limitations. Even when both modalities are present, misalignments are common and require careful registration. For platforms that do not support the same-slide H&E, the use of adjacent sections introduces biological variation. These issues complicate integration of histology with HR-ST data. Recent advances in virtual staining have shown that generative models can generate realistic H&E images from label-free[15–20] or fluorescence-labeled modalities[21]. Yet, no method exists to directly translate HR-ST into H&E images, despite HR-ST's unique ability to capture molecular and spatial information beyond imaging.

To address this gap, we propose ST2HE, the first framework for generating virtual H&E directly from HR-ST data. Built on Pix2Pix Turbo[22], a one-step latent diffusion model for high-fidelity image-to-image translation (Fig. 1b), ST2HE takes 4′,6-diamidino-2-phenylindole (DAPI) image overlaid with locations of spatial transcripts as inputs, where the image helps learn nuclear morphology and transcript locations to describe spatial gene expression (Fig. 1c). Gene-specific signals are encoded as colored dots, with variable color assignment across image tiles to ensure independence from specific gene panels, thus improving generalizability across tissues and HR-ST platforms. We trained ST2HE using Xenium data spanning five tissue types (Fig. 1d) and developed two complementary variants: ST2HE-CondGen, a tissue-conditioned model for training tissue types and ST2HE-UnconGen, a tissue-independent model for unseen tissue contexts (Fig. 1e). Together, these two models provide flexibility for real-world applications and enable direct use of pretrained pathology models like CONCH[13] to annotate HR-ST samples. By bridging HR-ST with histopathology, ST2HE expands the utility of HR-ST for discovery, precision medicine, and translational research.

## Results

**ST2HE generates histologically accurate virtual H&E for Xenium breast cancer samples**

We evaluated ST2HE using a publicly available Xenium breast cancer dataset, which provides whole-tissue ST profiling of breast cancer samples (Methods). Cell types were previously annotated by integrating scRNA-seq and H&E imaging, identifying major populations, including two tumor subtypes: ductal carcinoma in situ (DCIS) and invasive carcinoma[23] (Fig. 2a). Although these subtypes occupy distinct regions in gene expression space (Fig. 2a), they remain challenging to annotate without histological context. Notably, spatial niche clustering grouped cells from both subtypes into the same cluster, showing that clustering without histopathology fail to distinguish them (Fig. 2b). This underscores the importance of generating histopathology information for accurate annotation of HR-ST data.

Because breast cancer is a training tissue types, we used the conditional ST2HE to generate virtual H&E images (Fig. 2c), enabling the generation of more accurate breast cancer-specific morphology. This dataset includes paired, ground truth H&E staining, thus allowing evaluation of the generated H&E via pixel-wise similarity, structural comparison, and tissue-level assessment. We first assessed the generated image quality using mean squared error (MSE), peak signal-to-noise ratio (PSNR), and structural similarity index (SSIM). Across the dataset, ST2HE achieved an average MSE of 928.18, PSNR of 18.60 dB, and SSIM of 0.59. Stromal regions showed the highest structural similarity (SSIM = 0.60), followed by DCIS (SSIM = 0.58) and invasive tumor regions (SSIM = 0.57), with all pairwise differences being statistically significant ([Tukey's Honestly Significant Difference](), $p < 0.05$). PSNR was also higher in stromal tissue compared with invasive tumor (PSNR = 18.67 vs 18.41 dB, $p = 0.0077$), while differences in MSE among tissue types were not significant ($p = 0.11$) (Fig. 2e). These results indicate that ST2HE best preserves structural features in stromal regions, with reduced performance in tumor areas, likely due to their complexity. We next evaluated sample-level perceptual similarity with the Fréchet Inception Distance (FID) and obtained a score of 83.1, indicating that the generated H&E capture key structural features. Unlike traditional virtual staining, our method relies solely on transcript locations and DAPI signals, making morphology reconstruction more challenging. Beyond quantitative metrics, we also analyzed structural fidelity. Nucleus positions and shapes were well preserved, guided by the DAPI signals and cell morphology and cytoplasmic regions were faithfully reconstructed from ST signals. The spatial arrangement of colored transcript signals also helped the model learn morphology distributions, thus enhancing overall pathological patterns (Fig. 2d). Overall, ST2HE preserves both pixel details and histological structures, producing virtual H&E images like real slides.

We then evaluated annotation capability by fine-tuning the CONCH's image encoder[13] using H&E tiles from the BRACS dataset[24], which contains annotations of multiple breast cancer subtypes. The classifier achieved an AUC of 0.83 for normal, 0.75 for pathological benign, 0.6 for usual ductal hyperplasia, 0.62 for atypical ductal hyperplasia, 0.8 for DCIS, and 0.91 for invasive carcinoma (Fig. 2f). These results validate its use for annotation prediction. We next applied the classifier to virtual H&E images and compared the predicted labels with those obtained from real H&E images (Fig. 2h). The annotation transfer performance was quantified using AUC for each class, with the virtual H&E achieving 0.78 for normal, 0.89 for pathological benign, 0.92 for usual ductal hyperplasia, 0.76 for atypical ductal hyperplasia, 0.82 for DCIS and 0.81 for invasive carcinoma. With a weighted average AUC of 0.83 for real H&E images, this demonstrates strong agreement in predicted annotations (Fig. 2g), confirming ST2HE's utility for downstream breast cancer analysis.

**ST2HE Enables Cross-Platform Virtual H&E Staining on NanoString CosMx Data**

To evaluate the cross-platform capability of our model, we tested ST2HE using the NanoString NSCLC dataset (see Methods). This dataset includes three representative lung samples (Lung5_rep1, Lung6, and Lung9_rep1) with HR-ST profiles annotated across multiple biological layers: cell types, niches, and tumor subtypes[25]. Tumor subtype annotations were performed by expert pathologists at the UPMC Hillman Cancer Center, using DAPI staining, spatial cell type distributions, and fluorescence-based imaging[26] (Fig. 3a). The primary challenge in this setting is that NanoString utilizes a gene panel distinct from Xenium plus the cross-platform batch noise, which is ideal for testing ST2HE's robustness across platforms. Additionally, since several tumor subtypes appear in a different sample, the model must preserve subtype-specific pathological features while mitigating sample-specific variability. The lack of paired ground truth H&E images in this dataset further complicates direct evaluation of the generated virtual H&E. Consequently, we assessed the generated image quality through downstream analyses and expert annotations.

The generated H&E images, produced by the ST2HE-CondGen model (Fig. 3b), closely resemble authentic H&E staining and capture structural details that align with known pathology features. For instance, the micropapillary subtype exhibits well-preserved morphological characteristics, with epithelial clusters forming rounded papillary structures. All three NSCLC subtypes, i.e., micropapillary, solid, and complex acinar retain their distinguishing structural patterns, as confirmed by pathologist annotations[27] (Fig. 3c). To further assess the structural fidelity of the generated images, we embedded them using the CONCH model. UMAP visualization of CONCH embedding showed clear separation for micropapillary samples, while solid and complex acinar samples appeared more closely clustered but still visibly separated (Fig. 3d). This proximity may stem from batch effects, as both subtypes originate from the same lung sample (Lung9), possibly influencing their representation (Fig. 3d).

To annotate the generated images and compared the subtype predictions to the known annotations, we leveraged the full CONCH model, which employs both the image and text encoders in a contrastive framework. Unlike previous analyses where only image embeddings were used for visualization, here the model compares image embeddings with text embeddings of descriptive prompts (e.g., 'An image of {micropapillary tumor}') in a shared latent space and assigns the label with the closest match (Fig. 3e). Although initial classification results yielded AUC scores of 0.51 for complex acinar, 0.64 for solid, and 0.83 for micropapillary (Fig. 3f), UMAP visualizations have revealed that the generated H&E images preserved distinct subtype features. This discrepancy suggests a domain mismatch between the generated images and the data distribution used to train CONCH. To bridge this gap, we fine-tuned the CONCH model using additional generated H&E images from consecutive NanoString slides (see Methods). This realignment significantly improved classification performance, resulting in AUC scores of 0.72 for complex acinar, 0.74 for solid, and 0.93 for micropapillary (Fig. 3f). These improvements underscore the effectiveness of our synthetic H&E images in downstream pathology tasks.

To further evaluate the biological relevance of the generated H&E images, we used CONCH to identify tumor regions. We applied a sliding window approach (see Methods) to the classifier across image tiles, generating tumor subtype probability heatmaps from normalized prediction scores (Fig. 3g). These heatmaps showed strong alignment with ground truth annotations, providing additional validation. Notably, even after fine-tuning, the CONCH model continued to focus specifically on tumor regions, indicating that ST2HE learned generalized morphological patterns rather than overfitting to the generated images. Had the synthetic H&E images lacked biological fidelity, CONCH would not consistently attend to tumor regions. Thus, this result confirms both the biological integrity of the generated images and ST2HE's suitability for automated pathological phenotype detection. Overall, our findings demonstrate ST2HE's ability to generalize beyond its training technology platform, successfully producing high-quality H&E images of NSCLC using HR-ST data from a different NanoString platform. Furthermore, the improved performance achieved by fine-tuning the classifier with synthetic images further demonstrate the quality and biological relevance of the generated images, confirming ST2HE's utility in pathology annotations for HR-ST.

**ST2HE Generalizes for Unseen Tissue Types in Kaposi's Sarcoma**

To evaluate the model's ability to generalize to an entirely unseen tissue type, we used the tissue-independent ST2HE-UncondGen model trained on Xenium datasets containing skin tissues (Fig. 4a). With a generic prompt {'dapi2he'}, we generated virtual H&E images for our in-house Xenium dataset of Kaposi's sarcoma (KS), a highly inflammatory tumor caused by infection with Kaposi's sarcoma-associated herpesvirus (KSHV)[28] (Fig. 4b) . The goal of this experiment was to assess whether ST2HE could generate histologically meaningful images for KS, a cancer type absent from the training set, and capture features sufficient to distinguish between different KS stages.

Our in-house KS dataset includes three slides of tissue arrays, each comprising 16, 16, 11 tissues cores of cutaneous KS lesions, respectively, derived from different KS patients (Fig. 4b). Each core is annotated with one of three clinically defined KS stages (Patch, Plaque, or Nodular) based on the H&E stains from adjacent slides[29]. The dataset captures rich spatial transcriptomics features. Using spatial neighboring cell type composition analysis, we defined eleven distinct tissue niches including differentiated and basal epidermis, immune niche, various stromal compartments such as vascular endothelial cells (VEC), transit-amplifying vascular endothelial cells (TA VEC), macrophages (MΦ), and T-cell stroma, and tumor niches including tumor boundary, tumor, and tumor core (Fig. 4c). Analysis of tissues across KS stages revealed a progression associated shift in niche composition, transitioning from predominantly epidermal and stromal niches toward tumor-associated niches as KS advances from Patch to Nodular stage (Fig. 4d). Notably, Nodular cores also contain niches of Patch and Plague stages, indicating pathology patterns in later stages overlap with those in earlier stages (Fig. 4c&d). This overlap may complicate the ability of a predictive model to accurately infer the KS stage.

We first examined the generated H&E images of the cores from these slides. Interestingly, visual inspection revealed that ST2HE captured general morphological differences across KS stages (Fig. 4e). Particularly, in the Patch cores, we observed sparsely distributed spindle cells, whereas the Plaque cores showed moderately increased spindle cell density with more irregular clustering and the Nodular cores exhibited highly cellular regions with tightly packed spindle cells forming more consolidated structures (Fig. 4e). These observations align with pathological characteristics of KS progression[29–31]. Although the ST2HE never encountered skin tissue during training, the generated images appeared to reflect structurally meaningful differences across disease stages.

To quantitatively evaluate these features, we embedded the generated H&E tiles using the CONCH image encoder and visualized the resulting latent space using UMAP (Fig. 4f). The embeddings revealed stage-associated clustering, with the Nodular cores forming a dominant, compact region and the Patch cores being separated more distinctly from the rest. The Plaque stage exhibits partial overlap with both (Fig. 4f, left panel). When colored by spatial niche, early-stage Patch regions predominantly aligned with the expected early-stage niches, showing relatively homogeneous composition. In contrast, late-stage Nodular regions exhibited extensive heterogeneity, encompassing nearly all spatial niches, while being heavily enriched for late-stage niches such as tumor core and tumor boundary (Fig. 4f, right panel). These patterns suggest that the model captured biologically meaningful differences reflective of KS stage and tumor architecture, despite it has never seen any KS data during training.

We next tested whether these virtual H&E images could support downstream classification tasks. A stage-specific classifier trained on real H&E tiles from consecutive sections was applied to the generated virtual H&E tiles. As expected, given the domain shift and the model's lack of exposure to skin-specific morphology during training, tile-level classification performance was moderate, with AUC scores of 0.68 for Patch, 0.71 for Plaque, and 0.84 for Nodular (Fig. 4g). These suboptimal results likely stem from the absence of skin tissue in the training dataset, which makes it challenging for the model to generate accurate H&E features for skin-derived KS samples and biological complexity in niche composition, where late-stage cores frequently include niches, thus morphological features associated with earlier stages. Since each tile covers only a limited spatial region, the histological context visible in the tile may not fully align with the core-level stage label. Given these challenges, we attempted a core-level classification by adopting a top-K pooling strategy (Fig. 4h). By aggregating predictions from the most confident tiles within a core, we substantially improved classification performance. Core-level AUCs rose to 0.97 for Patch, 0.91 for Plaque, and 0.92 for Nodular (Fig. 4g). These results demonstrate that while individual tiles may be noisier, ST2HE preserves essential morphological cues across multiple tiles to enable accurate core-level inference. In summary, these results underscore ST2HE's capacity to generalize to previously unseen tissue types. Even though the model was trained without any skin data, it was able to generate virtual H&E images for KS that reflect relevant biological patterns and support downstream classification. These findings indicate that spatial transcriptomics features learned from other tissues can encode broadly applicable histological signals, demonstrating the potential of virtual staining for use in diverse and novel tissue settings.

**Larger Context, Balanced Loss, and Multi-Colored Transcripts Improve Virtual H&E Generation**

A core innovation of ST2HE is integrating DAPI staining with transcript locations to generate virtual H&E images. To assess how input configurations affect output quality, we conducted ablation studies using the Xenium breast cancer data. We first examined the tile size, which determines local context. Models trained with 512×512 tiles outperformed those with 256×256 (AUC of 0.81 vs. 0.73; Table 1), demonstrating that larger context improves morphological fidelity and classification accuracy. Even larger tiles may offer further improvements, but memory constraints prevented us from exploring them in this study.

Next, we evaluated ST2HE's loss weighting, which balances contributions from global-context features and local structural details. Experiments showed that models prioritizing detailed morphology consistently performed better across tasks. Specifically, when the global-context loss was assigned to a higher weight (0.3, denoted as *hg*), performance declined compared to the default setting (0.1), with an average AUC of 0.72 versus 0.80 (Table 1). These results suggest that preserving fine-grained structural details is more critical for reliable interpretation and annotation than emphasizing global style.

We further compared conditional and non-conditional training. ST2HE-CondGen, which uses tissue prompt, slightly outperformed ST2HE-UnCondGen for downstream subtype classification (AUC 0.83 vs. 0.81; Table 1). However, ST2HE-CondGen depends on the availability of tissue-type labels, which are not always practical to obtain. To balance accuracy and flexibility, we retained both versions in our pipeline, applying ST2HE-CondGen when labels are available and ST2HE-UnCondGen when they are not.

An important design choice in ST2HE is how transcript locations are visualized within the input images. Each transcript is represented as a dot overlaid on the DAPI image with dot size controlled by the parameter $s$ in the Matplotlib scatter function, where a larger $s$ corresponds to a bigger dot. This parameter affects how gene expression information is perceived by the model. We systematically evaluated $s$ from 0.1 to 0.9 in increments of 0.2 using the LPIPS loss, which measured perceptual similarity, L2 loss, which quantified pixel-level alignment, the CLIP similarity, which assessed semantic correspondence between transcripts and generated H&E, and clean FID, which captured overall perceptual realism. These losses provide a more detailed view of dot size effects than downstream cancer subtype classification AUC alone, which, while useful, is less diagnostic. Results showed that LPIPS trends were similar for all dot sizes (Fig. 5a). L2 and CLIP similarity were highest for $s$ = 0.1, suggesting improved pixel-level and semantic alignment, while clean FID indicated that $s$ = 0.3 produced the most realistic H&E-like images (Fig. 5a). Classification AUC showed best performance at $s$ = 0.5, followed closely by $s$ = 0.3 (Fig. 5b). Larger dot sizes caused transcript overlap, reducing spatial resolution and generating blurrier, less distinct features (Fig. 5c). Balancing these outcomes, we selected $s$ = 0.3 as the optimal dot size for all experiments, as it preserves transcript visibility while maintaining nuclear morphology.

Finally, we compared three visualization strategies: multi-colored transcripts (each gene assigned a unique color), mono-colored transcripts (all genes shown in a single color), and a DAPI-only input with no transcript visualization. Multi-colored transcripts consistently performed the best (AUC of 0.83 vs. 0.75 for mono-colored and 0.64 for DAPI-only inputs; Fig. 5d). This result indicates the multi-colored approach provides a rich visual cue that helps the model distinguish spatial gene expression patterns and associate them with morphological context. In contrast, mono-colored inputs led the model to confuse transcript dots with DAPI nuclear signals, producing scattered, small cell-like artifacts rather than coherent tissue morphology (Fig. 5e). DAPI-only inputs lacked transcript context and produced poor image quality (Fig. 5e). These findings support multi-colored transcript visualization as the default configuration.

**Discussion**

We developed ST2HE, a generative framework that synthesizes histologically realistic virtual H&E images directly from HR-ST data. By leveraging DAPI morphology and gene-level spatial context, ST2HE bridges the gap between transcriptomics and histopathology. Across datasets including breast cancer, NSCLC, and KS, generated images preserved key morphological features and supported downstream tasks such as subtype classification and stage prediction, demonstrating the feasibility of virtual H&E for spatial biology and computational pathology.

A major strength of ST2HE is its flexibility across platforms and tissues, enabled by dual conditional and generalization modes. In cross-platform and unseen-tissue evaluations, ST2HE produced images resembling real H&E while retaining predictive accuracy with pretrained models like CONCH, underscoring the fidelity of integrating transcriptional signals with generative modeling.

ST2HE also provides interpretability and modularity. Ablation studies showed that smaller transcript dots and multi-colored encodings improved structural extraction and preserved spatial context. Overlaying transcript locations on DAPI images enhanced interpretability, linking spatial gene expression to tissue morphology. These findings emphasize the importance of input design for maximizing fidelity and functional utility in HR-ST to H&E translation.

Some limitations remain. Current image quality, while suitable for computational use, lacks the fine-grained resolution required for clinical diagnostics. Future improvements could integrate additional imaging modalities (e.g., membrane stains or multiplex fluorescence) to capture finer detail and enable clinical interpretability. Beyond this, ST2HE could underpin spatial annotation databases, where virtual stains help transfer expert or model-derived annotations to HR-ST data, enriching systematic analyses of spatial gene expression and advancing disease classification and biomarker discovery.

**Method**

**Data preparation**

Raw spatial transcriptomics data and their corresponding tissue images were initially curated and processed to create suitable inputs for model training. The workflow involved three key steps: (1) aligning DAPI and H&E images through image registration, (2) filtering and overlaying transcript coordinates onto DAPI images, and (3) generating transcript-enhanced DAPI/H&E patch pairs. These processed pairs are the input and output for model training. We utilized five Xenium FFPE datasets from 10x Genomics: 1) preview FFPE Human Breast Sample, 2) Human Lung Cancer with Human Immuno-Oncology Profiling Panel and Custom Add-on, 3) FFPE Human Pancreatic Ductal Adenocarcinoma with Human Immuno-Oncology Profiling Panel, 4) FFPE Human Brain Cancer with Human Immuno-Oncology Profiling Panel and Custom Add-on, and 5) Human Liver with the Xenium Human Multi-Tissue and Cancer Panel, as Xenium provides both high-resolution DAPI images with transcript coordinates and matched post-Xenium H&E images. This pairing makes them uniquely suited for supervised learning, since ground-truth stained images are available for direct comparison.

**Image Registration**

Although the DAPI and post-Xenium H&E images originate from the same tissue slide, they cannot be perfectly identical because they are captured through different imaging processes. Small shifts, rotations, and scale differences arise during imaging and sample handling, which requires image registration for pixel level paired image translation. Spatial transcriptomic coordinates, however, do not require separate registration because they are intrinsically mapped onto the DAPI image during the Xenium assay. Once the DAPI and H&E images are aligned, the spatial transcriptomics are naturally aligned as well. For each tissue sample, the DAPI and H&E images were first loaded using a multi-resolution OME-TIFF reader. Maximum intensity projection was applied to the DAPI stack (except for breast tissue) to convert it into a 2D image. Each dataset includes a CSV file containing a 3×3 transformation matrix for image alignment. Both DAPI and H&E images were resampled to a common resolution of 0.2125 µm/pixel using bicubic interpolation. To ensure geometric consistency, we applied the transformation matrix within the rescaled coordinate frame using linear perspective warping. This resulted in aligned DAPI and H&E images with pixel-level correspondence, enabling paired training of the virtual staining model.

**Model training data generation**

To prepare image tiles for model training, we first selected representative regions across the tissue. Farthest point sampling (FPS) was applied to segmented cell centroids to identify spatially diverse index cells. For each selected cell, a 512×512 pixel window was extracted from the DAPI image and the corresponding post-Xenium H&E image, provided the tile contained sufficient content. Tiles with over 90% black or near-white pixels (grayscale 0 or >230) were discarded. DAPI tiles were normalized and saved as grayscale PNGs, and H&E tiles were saved in RGB format. Transcriptomic coordinates with quality scores above 20 were retained for overlay. Transcripts were plotted on the DAPI background as colored dots, with each gene assigned a unique color from a hue-sorted palette. The palette was independently randomized per tile to prevent color-gene correspondence across tiles, and dot sizes were varied to improve model robustness. Axes, labels, and spines were removed to produce clean overlays. The resulting transcript-enhanced DAPI images, comprising 2,000 tiles per tissue type, were exported and split into training and validation sets for model development and evaluation.

**ST2HE Model for Virtual H&E Staining**

ST2HE was built based on the Pix2Pix-Turbo model, a one-step image-to-image translation model designed for paired image synthesis. Pix2Pix-Turbo builds upon the traditional Pix2Pix framework by integrating a single-step diffusion-based generator with adversarial learning objectives, significantly reducing inference time while maintaining image quality. Additionally, it incorporated text-based conditional control to guide the generation of different tissue types, utilizing a CLIP-based text-image alignment loss.

The ST2HE generator is based on a pre-trained one-step diffusion model, which has been adapted for paired image-to-image translation. It employs LoRA (Low-Rank Adaptation) to efficiently fine-tune the model for virtual staining while reducing computational overhead. The network structure consolidates encoder, U-Net, and decoder components into an end-to-end trainable model with skip connections via zero-convolution layers to preserve fine image details (Fig. 1C). Text embeddings were processed through a CLIP encoder and integrated into the generator to modulate the output according to specified tissue types. The ST2HE discriminator is based on CLIP backbone, following the Vision-Aided GAN approach[32]. Since CLIP provides both text and image embeddings, the image embeddings are used for adversarial training, ensuring that the generated images align semantically with real stained images. This enables the model to enforce realism beyond tile-based discrimination, making it more suitable for text-guided virtual staining.

Mathematically, the loss function of the Pix2Pix-Turbo model consists of three terms:

1. **Adversarial Loss:**

$$\mathcal{L}_{GAN} = E_y \left[ \log D \left( E_{img}(y) \right) \right] + E_x \left[ \log \left( 1 - D \left( E_{img}(G(x,t)) \right) \right) \right] \quad (1)$$

where $G$ is the generator, $D$ is the discriminator operating on CLIP image embeddings $E_{img}(\cdot)$, $x$ is the input unstained image, $y$ is the corresponding H&E-stained image, and $t$ is the text conditioning input.

2. **Reconstruction Loss:**

$$\mathcal{L}_{rec} = E_{x,y}[|y - G(x,t)|_2] + LPIPS(G(x,t), y) \quad (2)$$

This term ensures that the generated images remain structurally similar to the ground truth using both L2 and LPIPS perceptual loss.

3. **CLIP Loss:**

$$\mathcal{L}_{CLIP} = -E_{x,t} \left[ \text{cosine\_similarity} \left( E_{img}(G(x,t)), E_{txt}(t) \right) \right] \quad (3)$$

where $E_{txt}(\cdot)$ represents the CLIP text embeddings, ensuring semantic alignment between the generated image and the specified tissue type.

The total loss function is given by:

$$G^* = \arg \min_G \max_D \mathcal{L}_{rec} + \lambda_{GAN}\mathcal{L}_{GAN} + \lambda_{CLIP}\mathcal{L}_{CLIP} \quad (4)$$

where $\lambda_{GAN}$ and $\lambda_{CLIP}$ are weighting factors that balance adversarial learning, pixel-level reconstruction, and text-image alignment.

**Model Training**

Training the ST2HE model leverages paired DAPI-H&E patches and, optionally, tissue-specific prompts to guide the model. The network was trained to minimize the loss function equation (4) so that it preserved fine-grained morphological features, maintained global tissue context, and ensured semantic consistency with the target tissue type.

We also implemented a JSON configuration system to manage training prompts. In the conditional setting, each training input was paired with a tissue-specific prompt (e.g., "This is a {tissue} H&E image"). For the unconditional setting, all samples shared a generic prompt ("dapi2he"). These text prompts were embedded via CLIP and passed to the model for condition generation on the desired tissue type.

**Downstream analysis**

To evaluate the utility of the generated virtual H&E tiles, we assessed their suitability for downstream histopathological analyses across three distinct datasets representing different tissue types and pathological contexts. Specifically, we tested whether the generated images could support tumor subtype annotations using pathology foundation models.

**Pathology Foundation Model** In this work, we leveraged the CONCH (Contrastively Aligned Visual-Language Pretraining) foundation model to enable a broad spectrum of downstream histopathological analyses. CONCH, pretrained on extensive H&E slide datasets with paired textual annotations, learns generalizable visual-language representations that are well-suited for interpreting diverse pathological features. Its rich embeddings capture essential aspects of tissue architecture, cellular morphology, and disease-specific cues, making it a powerful starting point for diagnostic modeling.

**Breast Cancer Subtype Classification**. Generated H&E tiles were first produced from DAPI images overlaid with spatial transcript coordinates, following the same input preparation used during model training. These virtual H&E tiles served as input to the downstream classification workflow. For subtype classification, the CONCH image encoder was fine-tuned using the BRACS dataset, and a single fully connected layer was appended to the encoder output to predict diagnostic subtypes, including benign, atypical hyperplasia, ductal carcinoma in situ (DCIS), and invasive carcinoma. This approach allows the model to leverage the fine-grained morphological information captured in the generated H&E tiles for accurate subtype annotation.

**NSCLC Subtype Classification and Tumor Segmentation** Virtual H&E tiles were generated from DAPI images with spatial transcript overlays using the same input preparation as in training. The NanoString NSCLC dataset provides multiple consecutive DAPI and transcript stacks per sample. For testing, we used the center stack, while the surrounding consecutive stacks were used to generate virtual H&E images for model fine-tuning. Pathologist-provided subtype annotations were embedded as textual prompts, and the CONCH image encoder was fine-tuned on the generated H&E tiles paired with these text prompts. Subtype classification for the test stack was then performed by computing the similarity between generated H&E embeddings and candidate textual prompts representing each NSCLC subtype.

To visualize tumor localization, we applied a 64×64 kernel sliding over the generated H&E images with a stride of 16 pixels. Each kernel patch was scored by its embedding similarity to the tumor text prompt. The resulting scores were stacked and averaged across overlapping kernels to produce a normalized heatmap, which was then smoothed using a Gaussian filter to generate a continuous representation of tumor probability.

**KS stage tile and core classification** for the KS dataset, lesions were classified into Patch, Plaque, and Nodular stages, reflecting progressive angio-proliferative changes. Real H&E images from three consecutive slides per case, paired with pathologist annotations, were used to fine-tune the CONCH image encoder with an appended fully connected classification layer. For tile-level predictions, each H&E tile was classified individually, capturing local morphological features such as vascular patterns and spindle cell density. For core-level predictions, we aggregated tile-level scores using top-k pooling, assigning the KS stage of the top-scoring tiles to the entire tissue core, thus providing a robust core-level classification.

**Evaluation Data**

To assess the generalization and effectiveness of our approach, we curated evaluation datasets spanning three distinct pathological contexts.

**Breast Cancer**: Evaluation was conducted on the Preview FFPE Human Breast Sample 1, a Stage IIB HER2-positive breast cancer tissue containing both ductal carcinoma in situ (DCIS) and invasive carcinoma. The sample was analyzed using the Xenium In Situ platform on a 5 µm FFPE section and co-registered with post-Xenium H&E and immunofluorescence (IF) images to support spatially resolved tumor and microenvironment analysis.

**Non-Small Cell Lung Cancer:** Evaluation utilized NanoString slides lung5, lung6, and lung9, which represent multiple NSCLC tumor subtypes. Detailed pathological review identified complex acinar, micropapillary, and solid subtypes. Subtype annotations were provided with the assistance of a pathologist. These samples enabled robust testing of the model's ability to capture morphological diversity and spatial transcriptomic variation within lung tumors.

**Kaposi's Sarcoma**: Evaluation was performed using an in-house dataset generated with the Xenium platform (10x Genomics) on three slides of tissue microarrays, each comprising of 16, 16, 11 tissues cores of cutaneous KS lesions annotated with corresponding KS stages (Patch,

Plaque, and Nodular). The dataset was generated using the Xenium Human Skin Gene Expression Panel Kit and a custom designed panel including a total of 306 genes targeting key major skin and immune cell markers, inflammatory and angiogenic regulators, tumor-associated genes, and representative KSHV latent and lytic genes.

## Data availability

The Xenium datasets for training and evaluating the ST2HE model are available from the Xenium data resource webpage https://www.10xgenomics.com/datasets, the NanoString CosMx SMI NSCLC data are available from https://nanostring.com/products/cosmx-spatial-molecular-imager/ffpe-dataset/nsclc-ffpe-dataset/, and the breast carcinoma H&E dataset BRACS is available from https://www.bracs.icar.cnr.it/. Xenium KS data will be deposited to Zenodo and made available upon publication

## Author contributions

Z.L. conceived the study, developed the ST2HE framework, performed data acquisition, conducted analyses and visualization, and wrote the original draft. A.D. contributed to method development and data acquisition. W.M. performed data acquisition. Y.-C.C. contributed to writing (review and editing) and provided funding and resources. S.-J.G. contributed to writing (review and editing) and provided funding and resources. Y.H. conceived and supervised the study, contributed to method development, writing (original draft, review and editing), and provided funding and resources. All authors approved the final version of the manuscript.

## Code availability

All custom code, including the ST2HE model weights and scripts for input data preparation, will be publicly available. The repository will be fully documented and include scripts reproducing all analyses and figure panels. Package dependencies for training and inference will be provided to ensure exact reproducibility.

**Acknowledgements**

This study was supported by grants from the National Institutes of Health (U01CA279618 and R21GM155774 to Y. Huang; CA096512, CA284554, CA278812, CA291244 and CA124332 to S.-J. Gao; and R35GM154967 to Y.-C. Chiu) and UPMC Hillman Cancer Center Startup Funds to S.-J. Gao and Y. Huang, and in part by award P30CA047904. This research was also supported in part by the University of Pittsburgh Center for Research Computing and Data, RRID:SCR_022735. Specifically, this work used the HTC cluster, which is supported by S10OD028483.

**Ethics declarations**

Competing interests

The authors declare no competing interests.

## Figure 1

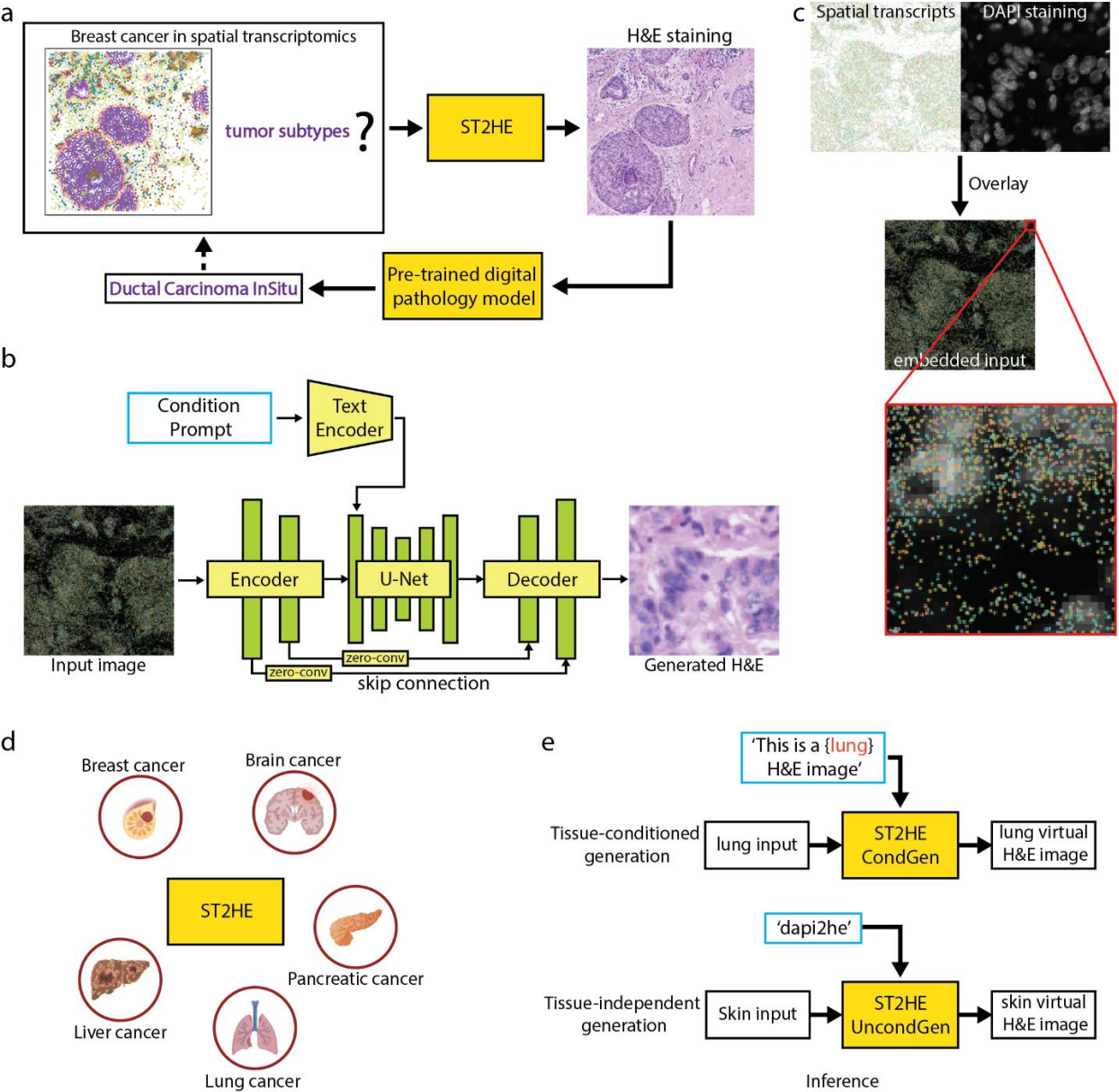

**Figure 1: Overview of the ST2HE framework.**

**a,** Challenge in annotating high-resolution spatial transcriptomics (HR-ST) data. Although HR-ST captures gene expression at single-cell resolution, tumor subtypes such as ductal carcinoma in situ (DCIS) remain difficult to assign directly from spatial profiles. In contrast, hematoxylin and eosin (H&E) staining is the clinical gold standard, supported by extensive digital pathology models.

**b,** ST2HE architecture for virtual H&E generation. Built on the Pix2Pix-Turbo backbone, the generator follows an encoder-U-Net-decoder design. Text prompts are encoded via CLIP and injected to guide tissue-specific virtual staining, producing images that align transcriptomic inputs with histological features.

**c,** Preparation of input tiles. DAPI images were sampled across tissues and overlaid with spatial transcript coordinates, represented as colored dots with randomized palettes to avoid fixed gene-color associations. These transcript-enhanced inputs form the basis for virtual H&E generation.

**d,** Training across multiple tissues. ST2HE was trained on Xenium data from five tissue types, capturing diverse morphologies and gene panels to support cross-platform generalization.

**e,** Variants of ST2HE. The tissue-conditioned model (ST2HE-CondGen) generates virtual H&E guided by tissue-specific prompts (e.g., "This is a lung H&E image"), while the tissue-independent model (ST2HE-UnconGen) uses a generic prompt ("dapi2he") to enable generation for tissues not represented in training.

Figure 2

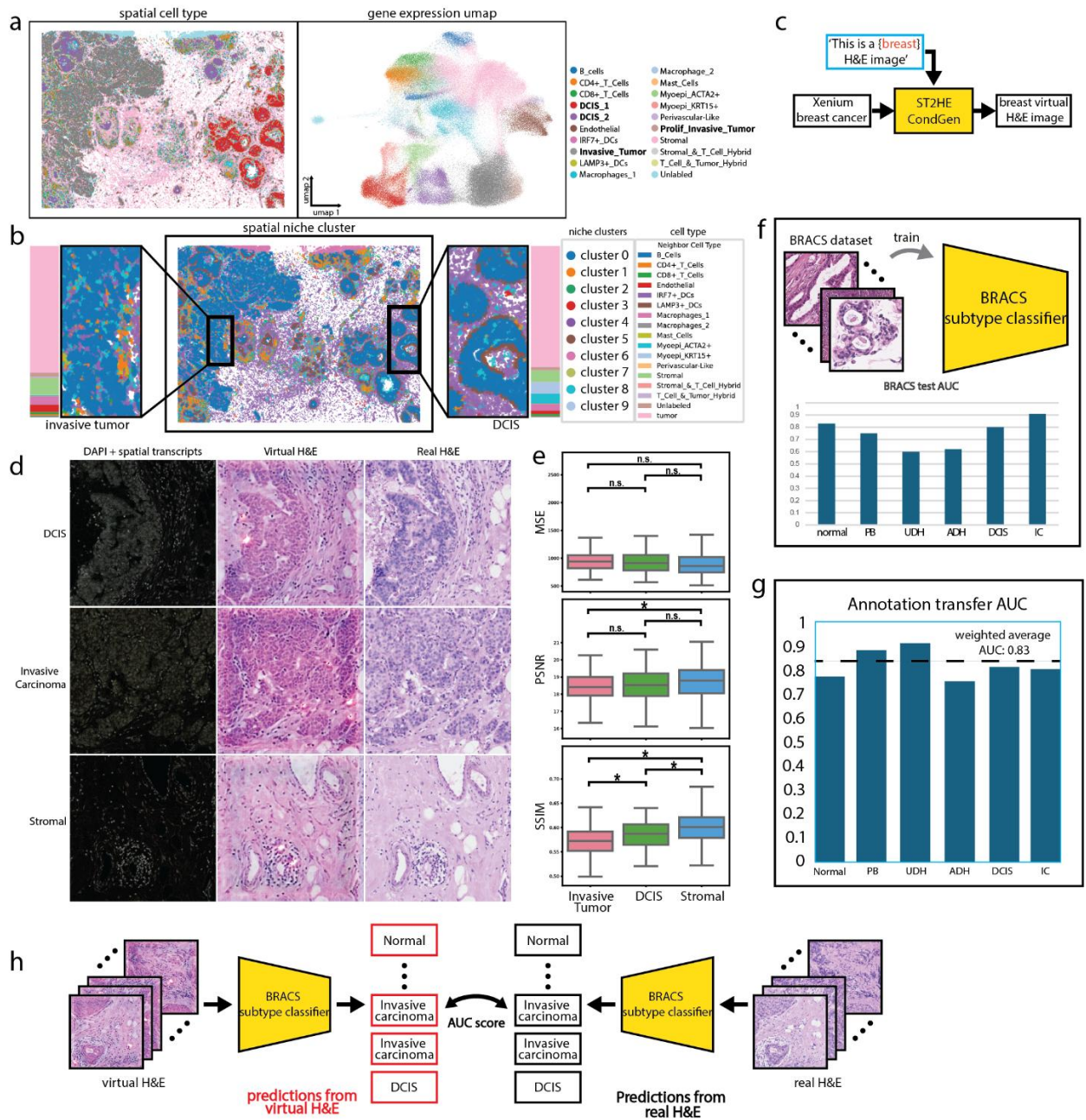

**Figure 2: Application of ST2HE to breast cancer spatial transcriptomics.**

**a,** Cell type visualization of a breast cancer sample. Left: spatial cell type map; right: UMAP embedding of gene expression colored by annotated cell types. Tumor-associated populations are grouped into ductal carcinoma in situ (DCIS) and invasive carcinoma (invasive tumor), with additional immune and stromal populations highlighting tissue heterogeneity.

**b,** Spatial niche clustering based on neighbouring cell composition. Cells from DCIS and invasive carcinoma were assigned to the same cluster, showing that clustering without histopathology cannot distinguish these tumor subtypes.

**c,** Virtual H&E images of breast cancer tissue generated by ST2HE-CondGen, leveraging breast cancer as a training tissue type.

**d,** Structural fidelity of virtual H&E images. Left: DAPI with transcriptomic coordinates (input); middle: virtual H&E generated by ST2HE; right: real H&E. Nuclei positions, morphology, and tissue architecture are faithfully reconstructed.

**e,** Pixel-level similarity between virtual and real H&E images across tissue types. Boxplots summarize MSE, PSNR and SSIM. Differences among tissue types were assessed by ANOVA; significant comparisons are marked with * and non-significant with n.s.

**f,** Performance of a breast cancer pathology classifier on real H&E images from the BRACS dataset. Bar plots show AUC values for six diagnostic categories: normal, pathological benign (PB), usual ductal hyperplasia (UDH), atypical ductal hyperplasia (ADH), DCIS, and invasive carcinoma.

**g,** Annotation transfer to virtual H&E images. Bar plots show AUC values for the same six categories, with performance comparable to real H&E, indicating preservation of pathological features.

**h,** Scheme of classification evaluation. A classifier fine-tuned on BRACS real H&E tiles was applied to virtual H&E images and predicted labels were compared with real H&E to assess annotation transfer.

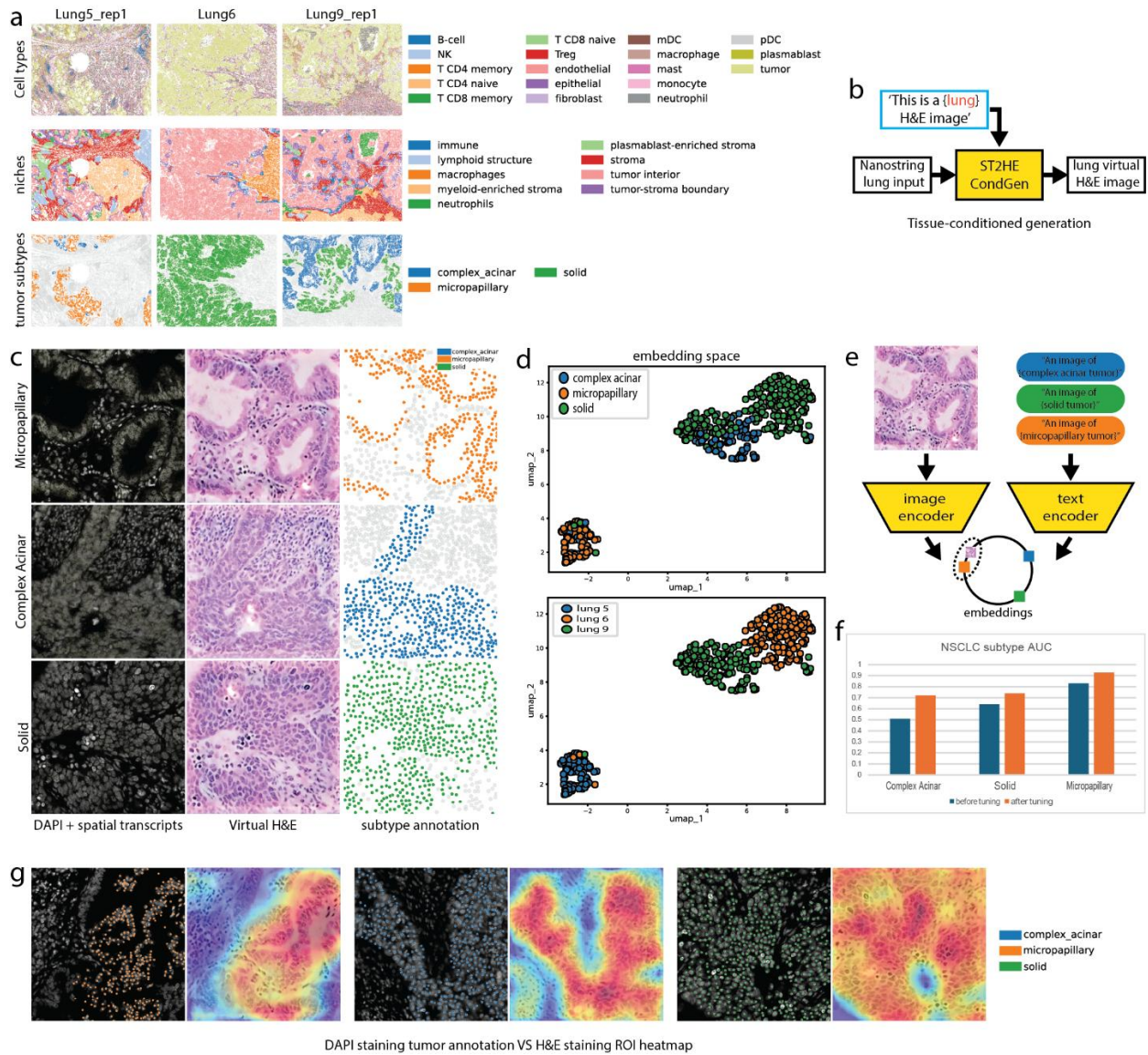

**Figure 3: Cross-platform application of ST2HE to non-small cell lung cancer.**

**a,** Spatial transcriptomics of three NanoString lung samples (Lung5_rep1, Lung6, Lung9_rep1). Cell types (top), spatial niches (middle), and tumor subtypes (bottom) were annotated by expert pathologists using DAPI and fluorescence images. Subtypes include complex acinar, micropapillary, and solid.

**b,** Tissue-conditioned generation of virtual H&E images using ST2HE-CondGen guided by lung-specific prompts.

**c,** Representative examples of three tumor subtypes. Left: DAPI with spatial transcripts; middle: virtual H&E generated by ST2HE; right: pathologist subtype annotations. Characteristic features such as papillary structures in micropapillary subtype are preserved.

**d,** UMAP embedding of CONCH image features from virtual H&E images. Top: subtype clusters colored by tumor subtype (complex acinar, micropapillary, solid). Bottom: the same embedding colored by sample identity (Lung5, Lung6, Lung9).

**e,** Schematic of zero-shot classification using CONCH. Virtual H&E images are embedded by the image encoder, while class-specific prompts (e.g., "An image of micropapillary tumor") are embedded by the text encoder. Both are projected into a shared contrastive space, and the image is assigned the label corresponding to the nearest text embedding.

**f,** NSCLC subtype classification performance. Initial AUC scores show variability across subtypes but fine-tuning CONCH with synthetic H&E images improves accuracy, particularly for complex acinar and solid subtypes.

**g,** Tumor region heatmaps from subtype probability scores. For each subtype, sliding-window predictions from the CONCH classifier are visualized as heatmaps overlaid on DAPI and virtual H&E images. Examples are shown for micropapillary, complex acinar, and solid subtypes, where predicted tumor regions align with ground truth annotations, confirming the biological relevance of ST2HE-generated H&E images.

Figure 4

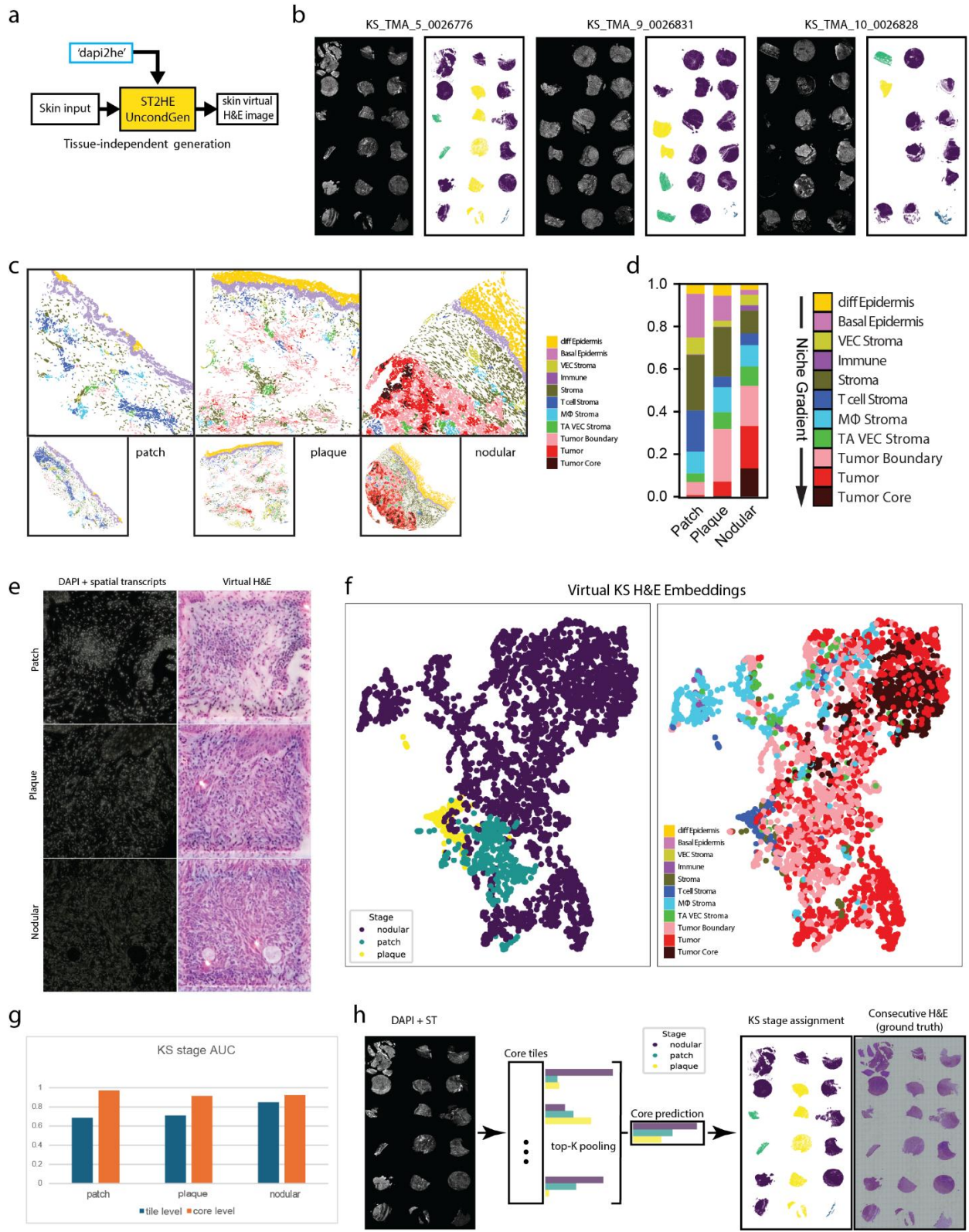

**Figure 4: Generalization of ST2HE to Kaposi's sarcoma, an unseen tissue type.**

**a,** Tissue-independent generation with ST2HE-UncondGen. A generic prompt ("dapi2he") was used to create virtual H&E images of KS, a cancer absent from the training set.

**b,** Representative DAPI inputs and virtual H&E images from three tissue microarray slides (KS_TMA_5, KS_TMA_9, KS_TMA_10). Each core is annotated with one of three KS stages: Patch, Plaque, or Nodular.

**c,** Spatial niche composition across KS stages. Eleven distinct niches, including epidermal, immune, stromal, and tumor-associated regions, were identified from spatial transcriptomics data.

**d,** Niche composition gradient across KS progression, showing a shift from epidermal/stromal niches in early patch lesions toward tumor and tumor core niches in Nodular lesions.

**e,** Generated virtual H&E images reflect stage-specific morphology. Patch lesions show sparse spindle cells, Plaque lesions exhibit increased density with irregular clustering, and Nodular lesions display tightly packed spindle cells forming consolidated structures.

**f,** UMAP embeddings of virtual H&E tiles. Left: clustering by KS stage. Right: clustering by spatial niche, showing heterogeneity in Nodular lesions and enrichment for tumor-associated niches.

**g,** KS stage classification performance. Bar plots show AUC for Patch, Plaque, and Nodular stages. Tile-level prediction performance is moderate but core-level performance improves markedly with top-K pooling.

**h,** Schematic of the top-K pooling strategy for core-level classification. Tile-level predictions from virtual H&E images are ranked by confidence and the most informative tiles are aggregated to generate a core-level prediction. This approach improves stage classification performance compared with tile-level evaluation.

Figure 5

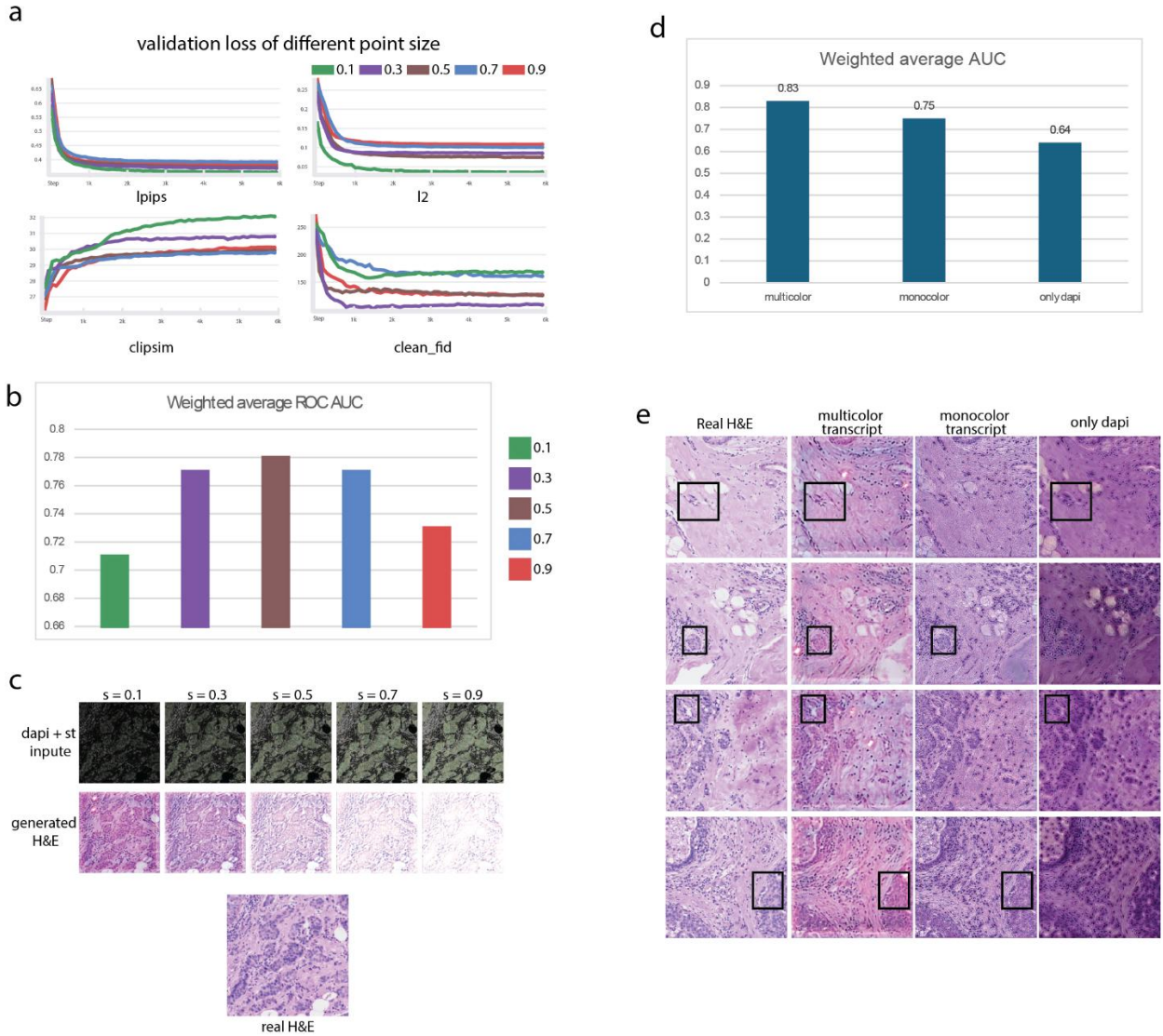

**Figure 5: Ablation study of context size, training loss, and transcript color scheme on virtual H&E generation.**

**a,** Validation performance of different transcript dot sizes ($s$ = 0.1-0.9) using LPIPS, L2, CLIP similarity, and clean FID, capturing perceptual, pixel-level, semantic, and realism-based quality.

**b,** Weighted average ROC AUC across dot sizes. Models performed best at intermediate dot sizes ($s$ = 0.3-0.7).

**c,** Representative virtual H&E images generated using different dot sizes. Larger dot sizes caused transcript overlap and blurrier morphology, while smaller sizes preserved nuclear features but reduced perceptual realism. Real H&E is shown for comparison.

**d,** Weighted average classification AUC for different transcript visualization strategies. Multi-colored transcript inputs outperformed mono-colored and DAPI-only inputs, demonstrating the importance of gene-specific color information.

**e,** Representative outputs under different input settings. Real H&E (left) is compared with virtual H&E generated by multi-colored transcripts, mono-colored transcripts, and DAPI-only inputs. Black boxes highlight regions where multi-colored inputs preserve structural fidelity and pathology-relevant features better than the alternatives.

**Table 1**. **Classification AUC performance of different ST2HE model variants across histopathological categories.**

Values represent classification AUC for each subtype. 'ST' and 'MT' indicate single-tissue and multi-tissue training, respectively, the subscripts (256, 512) denote tile size, '*hg*' indicates high weight of global context loss, and 'cond' indicates the use of conditional generation.

|        | ST_256_hg | MT_256 | MT_256_hg | ST_512 | MT_512 | MT_512_cond |
|--------|-----------|--------|-----------|--------|--------|-------------|
| Normal | 0.55      | 0.7    | 0.69      | 0.89   | **0.89** | 0.78      |
| PB     | 0.73      | 0.77   | 0.81      | **0.91** | 0.90 | 0.89        |
| UDH    | 0.75      | 0.86   | 0.80      | 0.92   | 0.91   | **0.92**    |
| ADH    | 0.83      | 0.82   | 0.80      | 0.78   | **0.85** | 0.76      |
| DCIS   | 0.79      | **0.86** | 0.81    | 0.79   | 0.73   | 0.82        |
| IC     | 0.63      | 0.74   | 0.72      | 0.76   | 0.80   | **0.81**    |
| avg    | 0.67      | 0.77   | 0.76      | 0.80   | 0.81   | **0.83**    |